\documentclass[journal, twoside, web]{ieeecolor}
\usepackage{generic}
\usepackage[english]{babel}

\newif\ifarxiv
\arxivfalse   

\usepackage{xcolor}
\let\labelindent\relax

\usepackage{amsmath,amssymb,amsthm}
\usepackage{enumitem}
\usepackage{mathrsfs}
\usepackage{graphicx}
\usepackage{caption,subcaption}    
\usepackage{setspace}
\usepackage{tipa}
\usepackage{booktabs}
\usepackage{newtxtext}
\usepackage{float}
\usepackage{epsfig}
\usepackage{epstopdf}
\usepackage{array}
\usepackage{algorithm}
\usepackage{algorithmicx}
\usepackage{algpseudocode}
\usepackage{textcomp}
\usepackage{threeparttable}

\usepackage{cite}
\usepackage[colorlinks,linkcolor=blue]{hyperref}
\usepackage[nameinlink,capitalize]{cleveref}
\usepackage{orcidlink}

\theoremstyle{plain}
\newtheorem{proposition}{Proposition}
\newtheorem{theorem}{Theorem}
\newtheorem{lemma}{Lemma}
\newtheorem{corollary}{Corollary}

\theoremstyle{definition}
\newtheorem{definition}{Definition}
\newtheorem{problem}{Problem}
\newtheorem{assumption}{Assumption}

\def\BibTeX{{\rm B\kern-.05em{\sc i\kern-.025em b}\kern-.08em
    T\kern-.1667em\lower.7ex\hbox{E}\kern-.125emX}}
\begin{document}
\title{Energy-Aware Distributed Task Allocation via Local Submodular Maximization}

\author{Jing~Liu,
Fangfei~Li,~\IEEEmembership{Member,~IEEE}, Xin~Jin,~\IEEEmembership{Member,~IEEE},
Yang~Tang,~\IEEEmembership{Fellow,~IEEE},
Ruggero~Carli,~\IEEEmembership{Member,~IEEE}, and~ Luca~Ballotta,~\IEEEmembership{Member,~IEEE}
\thanks{This work is supported by the Program of China Scholarship Council under Grant 202506740033, the National Key R\&D Program of China under Grant 2025YFA1016504, the National Natural Science Foundation of China under Grants 62233005, 62573198, U2441245, U25B6002, 62403141, the National Key Laboratory of Space Target Awareness under Grant STA2025ZCB0208, in part by the Shanghai Institute for Mathematics and Interdisciplinary Sciences (SIMIS) under Grant SIMIS-ID-2025-SP. \textit{(Corresponding authors: Fangfei Li and Yang Tang.)}}
\thanks{Jing Liu is with the School of Mathematics, East China University of Science and Technology, Shanghai 200237, China (e-mail: y20220091@mail.ecust.edu.cn).}
\thanks{Fangfei Li is with the School of Mathematics and the Key Laboratory of Smart Manufacturing in Energy Chemical Process, Ministry of Education, East China University of Science and Technology, Shanghai 200237, China (e-mail: li\_fangfei@163.com, lifangfei@ecust.edu.cn).}
\thanks{Xin Jin is with the Research Institute of Intelligent Complex Systems, Fudan University, Shanghai 200433, China (e-mail: jx\_9810@163.com).}
\thanks{Yang Tang is with the Key Laboratory of Smart Manufacturing in Energy Chemical Process, Ministry of Education, East China University of Science and Technology, Shanghai, 200237, China (e-mail: yangtang@ecust.edu.cn).}
\thanks{Ruggero Carli and Luca Ballotta are with the Department of Information Engineering, University of Padova, 35131 Padova, Italy (e-mail: carlirug@dei.unipd.it; luca.ballotta@unipd.it).}}

\maketitle

\begin{abstract}
This paper studies distributed task allocation under time-varying agent sets, task sets, and communication graphs. At each decision epoch, the stage allocation problem is formulated as monotone submodular maximization over a time-varying, energy-filtered admissible ground set subject to a partition-matroid constraint. We develop a Distributed Adaptive Greedy Bidding Algorithm (DAGBA) that combines local admissible-set construction, conditional marginal bidding, and neighbor-based conflict resolution under limited communication. We prove that DAGBA guarantees per-epoch feasibility, finite-round convergence, and residual-energy safety. Theoretically, we identify a sequential-greedy regime where DAGBA recovers the classical $1/2$-approximation guarantee. For general executions with partial information, we derive a local-information-factor approximation bound and conservative bounds computable from the execution trace that quantify the degradation caused by local estimates and parallel fixing. Simulations on cooperative observation missions demonstrate that DAGBA achieves task-completion progress and utility close to centralized and complete-communication references, while improving communication efficiency and maintaining strict energy feasibility.
\end{abstract}



\section{Introduction}
\label{sec.1}

The rapid deployment of autonomous agents in open environments, such as swarms of unmanned aerial vehicles, mobile sensor networks, and robotic search-and-rescue teams, has stimulated extensive research on adaptive and scalable task allocation strategies \cite{zhou2023robust,shi2023robust,chen2024computer,xue2024practical}. In these applications, agents must repeatedly coordinate to accomplish spatially and temporally evolving tasks, including target tracking, area coverage, and information gathering. In practice, the participating agent set, the active task set, and the communication topology may all change over time due to mobility, failures, energy depletion, and asynchronous task arrivals and completions \cite{liu2024distributed,tang2025resilient,jiang2025dynamic,wu2026open,deplano2026optimization,liu2025online,liu2025distributed,liu2026distributed}. These time-varying features challenge the assumptions underlying many classical distributed coordination methods, which typically require fixed agent teams, stable communication networks, and sufficient time for bidding or consensus to stabilize \cite{sarne2005solving,paccagnan2021utility,fu2022robust,wu2023predictive,choi2009consensus,liemhetcharat2012modeling,rogers2011bounded,Braquet2021GreedyDA}.

Recent studies on dynamic multi-agent coordination have made important progress in distributed optimization, online resource allocation, and open multi-agent systems (OMAS)~\cite{liu2024distributed,liu2025distributed,wu2026open,deplano2026optimization,liu2025online,liu2026distributed,tang2025resilient,jiang2025dynamic}. However, these works primarily address continuous-valued optimization problems with convex or linear objectives. The task allocation problem studied in this paper is combinatorial in nature. Its admissible decisions are discrete agent-task pairs, the utility is nonadditive due to cooperative sensing effects, and feasibility is coupled with local motion constraints and residual energy budgets. In particular, assigning additional agents to the same task often improves observation or tracking performance but with diminishing marginal benefit, a structure naturally formalized by monotone submodularity~\cite{qu2019distributed,krause2014SubmodularFM,liuj2026distributed}.

The present problem introduces several difficulties that are not simultaneously captured by existing distributed submodular maximization frameworks.
First, the admissible ground set is time-varying: at each decision epoch, candidate agent-task pairs depend on the currently active agents, active tasks, residual energies, and motion-feasibility conditions.
Second, each agent carries a finite onboard energy reserve over the mission horizon. Energy consumed at the current epoch reduces the residual energy available at subsequent epochs and may therefore restrict future admissible ground sets.
Third, the communication graph determines which bidding and confirmed-assignment information is locally visible to each agent. Under neighbor-limited communication, different agents may maintain different estimates of the accepted assignment, so the exact centralized greedy sequence used in classical matroid-constrained submodular maximization cannot in general be enforced.  Online optimization methods such as regret minimization, bandit allocation, and budgeted utility maximization \cite{badanidiyuru2018bandits,kapralov2013online,agrawal2014bandits} address submodular objectives in isolation, but they are not designed for this communication-limited, multi-agent setting with cross-epoch energy coupling. Existing distributed submodular maximization frameworks face the same limitation from a different angle: consensus-based bundle methods \cite{choi2009consensus,Braquet2021GreedyDA} and distributed sequential greedy algorithms \cite{qu2019distributed,gharesifard2017distributed,grimsman2018impact} typically assume a static agent set or a stable communication structure, while continuous-relaxation methods attaining the tight $(1-1/e)$-approximation of the multilinear extension \cite{calinescu2011maximizing,robey2021optimal} rely on oracle access and multiple rounds of gradient-consensus iterations per decision epoch, rendering them computationally prohibitive for per-epoch deployment in time-critical missions.

Motivated by these observations, this paper studies distributed task allocation under time-varying feasibility, residual-energy constraints, and local communication. At each decision epoch, the currently available information defines a stage allocation problem formulated as monotone submodular maximization over an energy-filtered admissible ground set subject to a partition-matroid constraint. Since future agent arrivals, task completions, and communication changes are unknown at decision time, we evaluate the proposed algorithm against the offline optimum of the same-epoch problem, which provides a natural and tractable per-epoch benchmark. To solve the stage problem, we develop the Distributed Adaptive Greedy Bidding Algorithm (DAGBA). Even against this per-epoch benchmark, establishing an approximation guarantee is nontrivial. Under neighbor-limited communication, each agent maintains only a local estimate of the accepted assignment and multiple pairs may be fixed in parallel, deviating from the exact sequential greedy order required by the classical $1/2$-approximation \cite{fisher1978analysis}.
Our contributions are as follows:
\begin{enumerate}
\item We formulate distributed task allocation in OMAS with time-varying agent sets, task sets, and communication graphs as a sequence of stagewise monotone submodular maximization problems over time-varying, energy-filtered admissible ground sets subject to partition-matroid constraints. Unlike recent dynamic multi-agent optimization methods that address continuous-valued convex objectives~\cite{deplano2026optimization,wu2026open,liu2025online,liu2026distributed}, this formulation simultaneously captures the combinatorial decision structure, cooperative diminishing returns, neighbor-limited communication, and cross-epoch coupling through residual energy constraints.

\item We propose DAGBA, a distributed greedy coordination algorithm based on local admissible-set construction, conditional marginal bidding, and neighbor-based conflict resolution. The central design challenge is to realize the greedy marginal-selection principle without a global coordinator: because the marginal gain of each candidate depends on which agents have already been assigned, every agent must bid against a locally maintained assignment estimate that is generally neither identical to nor synchronized with those of other agents. Unlike consensus-based bundle methods~\cite{choi2009consensus,Braquet2021GreedyDA} that require a stable agent set for bidding to converge, DAGBA accommodates time-varying agent and task sets by reconstructing local admissible subsets and local assignment estimates at each decision epoch, without requiring global consensus.

\item We establish per-epoch feasibility, finite-round convergence, and residual-energy safety of DAGBA, and derive per-epoch approximation guarantees. In a sequential-greedy regime, DAGBA recovers the classical $1/2$-approximation with a curvature-dependent refinement. For general local-information executions, we derive a local-information-factor approximation bound and conservative bounds computable at runtime that quantify the degradation caused by stale local estimates and parallel fixing. The analysis is nontrivial because, under sparse communication, different agents evaluate marginal bids with respect to different local estimates and multiple pairs may be fixed in parallel, deviating from the exact sequential greedy order assumed by classical matroid-constrained submodular maximization analysis~\cite{fisher1978analysis}. The proof requires a new charging argument that verifies local estimates are subsets of the final assignment, establishes the submodular charging inequality for arbitrary, not necessarily nested, pairs of assignments, and exploits the partition-matroid structure to prevent double-charging.

\end{enumerate}

\emph{Notation}: Let $\mathbb{Z}_+$ and $\mathbb{R}_+$ denote the sets of positive integers and nonnegative real numbers, respectively. For a finite set $\mathcal{A}$, $2^{\mathcal{A}}$ denotes its power set and $|\mathcal{A}|$ denotes its cardinality. For two sets $\mathcal{A}$ and $\mathcal{B}$, $\mathcal{A}\setminus\mathcal{B}$ denotes the set difference. The indicator function is denoted by $\mathbf{1}_{{\cdot}}$, which equals one if the condition in braces is true and zero otherwise. For a set function $f$, the marginal gain of adding an element $e$ to a set $S$ is denoted by $\Delta f(e\mid S)=f(S\cup\{e\})-f(S)$. For an agent $i$, $\mathcal{N}_{i,t}$ denotes its neighbor set and $\mathcal{N}_{i,t}^{+}=\mathcal{N}_{i,t}\cup\{i\}$ denotes its closed neighborhood. The symbol $\arg\max$ denotes the set of maximizers.

The remainder of this paper is organized as follows. 
Section~\ref{sec.2} formulates the problem and reviews 
preliminaries. Section~\ref{sec.3} presents DAGBA. 
Section~\ref{sec.4} provides the theoretical analysis. 
Section~\ref{sec.5} reports simulations. 
Section~\ref{sec.6} concludes the paper.

\section{Task Allocation Problem Formulation}\label{sec.2}

We consider distributed task allocation in an OMAS over a finite horizon $t\in\{1,\ldots,T\}$, where $T\in\mathbb{Z}_{+}$. Let $\mathcal{V}= \{1,2,\ldots, N_{\max}\}$ and $\mathcal{M} = \{1,2,\ldots, M_{\max}\}$ denote the universal sets of all possible agents and tasks that may appear during the entire mission horizon, respectively. At epoch $t$, only the active subsets $\mathcal{V}_t\subseteq \mathcal{V}$ and $\mathcal{M}_t\subseteq \mathcal{M}$ participate in the allocation problem. The active communication topology is described by an undirected graph $\mathcal{G}_t=(\mathcal{V}_t,\mathcal{E}_t)$. 
Since agents may join or leave and tasks may arrive or be completed between consecutive decision epochs \cite{deplano2026optimization}, the tuple $(\mathcal{V}_t,\mathcal{M}_t,\mathcal{G}_t)$ is generally time-varying, which in turn makes the admissible ground set and feasible family dependent on the decision epoch.

\subsection{Preliminaries}
\label{subsec:preliminaries}

Let $\Omega$ be a finite ground set and let $f:2^\Omega\to\mathbb{R}$ be a set function. The function $f$ is normalized if $f(\varnothing)=0$, monotone nondecreasing if $f(A)\le f(B)$ for all $A\subseteq B\subseteq \Omega$, and submodular if $f(A\cup\{e\})-f(A)\ge f(B\cup\{e\})-f(B)$
holds for all $A\subseteq B\subseteq \Omega$ and all $e\in \Omega\setminus B$. The marginal gain of adding $e$ to $A$ is denoted by $\Delta_f(e\mid A)=f(A\cup\{e\})-f(A)$~\cite{krause2014SubmodularFM}.

\begin{definition}[Matroid \cite{hou2021robust}]
Let $\Omega$ be a finite ground set and let $\mathcal{I}\subseteq 2^\Omega$ be a family of subsets of $\Omega$. The set system $(\Omega,\mathcal{I})$ is called a matroid if:
\begin{enumerate}[label=(\roman*)]
    \item $\emptyset\in\mathcal{I}$;
    \item if $A\in\mathcal{I}$ and $B\subseteq A$, then $B\in\mathcal{I}$;
    \item if $A,B\in\mathcal{I}$ and $|A|<|B|$, then there exists $e\in B\setminus A$ such that
    $A\cup\{e\}\in\mathcal{I}$.
\end{enumerate}
The sets in $\mathcal{I}$ are called independent sets.
\end{definition}

\begin{definition}[Partition matroid \cite{hou2021robust}]  \label{def:partition_matroid}
Let $\{P_1,P_2,\dots,P_m\}$ be a partition of the ground set $\Omega$, namely,
$\bigcup_{i=1}^m P_i=\Omega,\
P_i\cap P_j=\emptyset,\ \forall i\neq j$.
Let $k_i\in\mathbb{Z}_{+}$ satisfy $k_i\le |P_i|$ for all $i=1,2,\dots,m$, and define $\mathcal{I}  = \left\{ X\subseteq \Omega \middle|  |X\cap P_i|\le k_i,\ \forall i=1,2,\dots,m \right\}$.
Then $(\Omega,\mathcal{I})$ is called a partition matroid.
\end{definition}

\begin{definition}[Total curvature \cite{conforti1984submodular}]
\label{def:curvature}
Let $f:2^\Omega\to\mathbb{R}_+$ be a normalized monotone submodular function. The total curvature of $f$ is
\begin{equation}
    c_f = 1 - \min_{e\in\Omega:\Delta_f(e\mid\varnothing)>0}
    \frac{\Delta_f(e\mid \Omega\setminus\{e\})}{\Delta_f(e\mid\varnothing)} \in[0,1].
    \label{eq:curvature_def}
\end{equation}
\end{definition}
When $c_f=0$, $f$ is modular; when $c_f=1$, $f$ exhibits the maximum diminishing-returns effect.

\subsection{Problem of Interest}

We formulate the per-epoch task-allocation problem. For each active agent $i\in \mathcal{V}_t$, let $x_{i,t}\in\mathbb{R}^{n_x}$ be its state and $u_{i,t}\in\mathbb{R}^{n_u}$ be its control input. The motion layer evolves according to
\begin{equation}
    x_{i,t+1}=f_i(x_{i,t},u_{i,t}).
    \label{eq:agent_dyn}
\end{equation}
For each active task $j\in \mathcal{M}_t$, let $y_{j,t}\in\mathbb{R}^{n_y}$ denote the task state, with dynamics
\begin{equation}
    y_{j,t+1}=h_j(y_{j,t})+\xi_{j,t}, \ \xi_{j,t}\sim\mathcal{N}(0,Q_{j,t}).
    \label{eq:target_dyn}
\end{equation}

At epoch $t$, the currently available information is $\mathcal{H}_t = \left(
\mathcal{V}_t,\mathcal{M}_t,\mathcal{G}_t,
\{x_{i,t},E_{i,t}\}_{i\in\mathcal{V}_t},
\{y_{j,t}\}_{j\in\mathcal{M}_t}
\right)$, where $E_{i,t}\ge0$ denotes the residual energy of agent $i$. 
The agent dynamics in \eqref{eq:agent_dyn} are modeled deterministically at the allocation layer, since the current agent state is estimated to be available and the nominal motion response is handled by the lower-level controller or planner. By contrast, task dynamics are not controlled by the agents and may be affected by exogenous disturbances.
Hence, the process noise $\xi_{j,t}$ is included in \eqref{eq:target_dyn} to model uncertainty in task evolution \cite{zhou2023robust}. 

An assignment at epoch $t$ is a subset $\Pi_t\subseteq \mathcal{V}_t\times \mathcal{M}_t$, where $(i,j)\in\Pi_t$ means that agent $i$ is assigned to task $j$ at epoch $t$. Since multiple agents may jointly serve the same task, we associate each assignment $\Pi$ with the subset of agents assigned to task $j$, defined by
$\mathcal{V}_{j,t}(\Pi)=\{i\in \mathcal{V}_t\mid (i,j)\in \Pi\}$.
The allocation utility at epoch $t$ is then given by
\begin{equation}
    U_t(\Pi)=\sum_{j\in \mathcal{M}_t}F_{j,t}\big(\mathcal{V}_{j,t}(\Pi)\big).
    \label{eq:stage_utility}
\end{equation}
Here, each $F_{j,t}:2^{\mathcal{V}_t}\to\mathbb{R}_{+}$ evaluates the utility obtained from the subset of agents assigned to task $j$ and is assumed to be normalized, monotone nondecreasing, and submodular. The additive structure in \eqref{eq:stage_utility} assumes task independence \cite{qu2019distributed}, while the submodularity of $F_{j,t}$ captures diminishing marginal returns caused by sensing overlap or redundancy~\cite{krause2014SubmodularFM,qu2019distributed}. Since $U_t$ is a sum of nonnegative submodular functions composed with disjoint agent subsets, $U_t$ is a normalized monotone submodular function over the agent-task pair ground set $\Omega_t$ \cite{krause2014SubmodularFM}.

For each candidate pair $(i,j)\in \mathcal{V}_t\times \mathcal{M}_t$, 
the lower-level planner returns a predicted execution cost 
$\widehat{C}_{ij,t}\in[0,+\infty]$, which estimates the energy required for agent $i$ to move toward and execute task $j$ along a locally feasible trajectory. The predicted execution cost $\widehat{C}_{ij,t}$ is computed by the lower-level motion planner based on the agent's current state $x_{i,t}$, the task's current state $y_{j,t}$, and the motion dynamics \eqref{eq:agent_dyn}.
If no such trajectory exists, we set $\widehat{C}_{ij,t}=+\infty$. 
The admissible ground set at epoch $t$ is defined as
\begin{equation}
    \Omega_t=\big\{(i,j)\in \mathcal{V}_t\times \mathcal{M}_t
    \mid \widehat{C}_{ij,t}\le E_{i,t}\big\}.
     \label{eq:agent-task}
\end{equation}
Thus, every pair in $\Omega_t$ is individually executable under the current residual-energy and local motion-planning conditions.
The feasible family at epoch $t$ is
\begin{equation}
    \mathcal{S}_t
    =
    \Big\{
    \Pi\subseteq \Omega_t
    \ \Big|\
    \sum_{j\in \mathcal{M}_t}\mathbf{1}_{(i,j)\in\Pi}\le 1,\ \forall i\in \mathcal{V}_t
    \Big\}.
    \label{eq:feasible_family}
\end{equation}
Hence each active agent can serve at most one task at epoch $t$, while multiple agents may be assigned to the same task.
This feasible family $\mathcal{S}_t$ forms a \textit{partition matroid} over the ground set $\Omega_t$, where each partition corresponds to all admissible task pairs of a single agent, and each partition can contain at most one selected element. For each epoch $t$, let $c_t$ denote the total curvature of the induced stage utility $U_t$ over $\Omega_t$. 
After the assignment is executed for one epoch, the residual energy evolves according to the realized one-step execution cost:
\begin{equation}
E_{i,t+1}=E_{i,t}-
\sum_{j\in\mathcal{M}_t}e_{ij,t}\mathbf{1}_{(i,j)\in\Pi_t},
\quad i\in\mathcal{V}_t,
\label{eq:energy_update}
\end{equation}
where $e_{ij,t}\ge 0$ denotes the energy actually consumed by agent $i$ for executing the current epoch of the planned motion toward task $j$. 

\begin{assumption}[Execution-cost prediction]
\label{ass:cost_upper_bound}
For every epoch $t$ and every admissible pair $(i,j)\in\Omega_t$, the predicted execution cost returned by the lower-level planner upper bounds the realized one-epoch execution cost, namely,
\begin{equation}
    0\le e_{ij,t}\le \widehat{C}_{ij,t},
    \ (i,j)\in\Omega_t .
    \label{eq:execution_cost_upper_bound}
\end{equation}
\end{assumption}
The energy constraint enters the per-epoch problem through the admissible ground set~\eqref{eq:agent-task}. Across epochs, energy consumed by the current assignment reduces $E_{i,t+1}$ in~\eqref{eq:energy_update}, which may reduce $\Omega_{t+1}$.

We consider the following stage allocation problem at epoch $t$.
\begin{problem}
\label{prob:stage_allocation}
Given the currently available information $\mathcal{H}_t$, determine an assignment $\Pi_t\in \mathcal{S}_t$ such that
\begin{equation}
    \max_{\Pi\in \mathcal{S}_t} U_t(\Pi).
    \tag{P1}
    \label{eq:stage_problem}
\end{equation}
\end{problem}
Problem~\ref{prob:stage_allocation} is solved at each epoch using only $\mathcal{H}_t$. Because future agent arrivals, task arrivals, departures, and communication changes are unknown at decision time, the performance analysis evaluates the quality of $\Pi_t$ against the offline optimum of the same-epoch problem. This per-epoch benchmark is natural and tractable, isolating the distributed coordination difficulty from the prediction of future system events.

\begin{assumption}[Local information and communication~\cite{qu2019distributed}]
\label{ass:local_information}
At each epoch $t\in\{1,\ldots,T\}$, the communication graph
$\mathcal{G}_t=(\mathcal{V}_t,\mathcal{E}_t)$ is undirected and need not be connected.
Each active agent $i\in\mathcal{V}_t$ knows its own state and residual energy
and has access to the predicted execution costs and task information required
to construct its local admissible set $\Omega_{i,t}$.
Agent $i$ communicates only with its current neighbors
$\mathcal{N}_{i,t}=\{k\in\mathcal{V}_t\mid(i,k)\in\mathcal{E}_t\}$
and has no access to the global assignment state or the full admissible
ground set $\Omega_t$.
For every locally admissible pair $(i,j)\in\Omega_{i,t}$ and every locally
available confirmed-assignment estimate $\widehat{\Pi}_{i,t}$, agent $i$ can
evaluate the conditional marginal gain
$\Delta_t\bigl((i,j)\mid\widehat{\Pi}_{i,t}\bigr)$ using the local assignment estimate $\widehat{\Pi}_{i,t}$ and the task utility parameters obtained through neighbor communication.
\end{assumption}


\section{Distributed Greedy Algorithm for Resource-Constrained Submodular Task Allocation}
\label{sec.3}

\begin{algorithm}[t]
\caption{DAGBA at epoch $t$}
\label{alg:dagba_main}
\begin{algorithmic}[1]
\State The current system state determines $\mathcal{V}_t$, $\mathcal{M}_t$, $\mathcal{G}_t$, and $\Omega_t$.
\State Each active agent $i\in\mathcal{V}_t$ constructs $\Omega_{i,t}$ in \eqref{eq:agent_subset}.
\State Each agent $i$ initializes $\mu_{i,k,t}^{(0)}\gets 0$, $\beta_{i,k,t}^{(0)}\gets 0$, and $\phi_{i,k,t}^{(0)}\gets 0$ for all $k\in\mathcal{V}_t$.
\For{$r=1,2,\ldots,R_t^{\max}$}
    \ForAll{$i\in\mathcal{V}_t$}
    \State Update previous local information state:
$
\mu_{i,t}^{(r)}\gets\mu_{i,t}^{(r-1)},\
\beta_{i,t}^{(r)}\gets\beta_{i,t}^{(r-1)},\
\phi_{i,t}^{(r)}\gets\phi_{i,t}^{(r-1)}.
$
        \State Construct the local assignment estimate
        $
        \widehat{\Pi}_{i,t}^{(r-1)}
        =
        \{(k,\mu_{i,k,t}^{(r-1)})
        \mid
        \phi_{i,k,t}^{(r-1)}=1,\ 
        \mu_{i,k,t}^{(r-1)}\neq 0\}.
        $
        \If{$\phi_{i,i,t}^{(r-1)}=0$ and $\Omega_{i,t}\neq\varnothing$}
            \State $(j_{i,t}^{(r)},b_{i,t}^{(r)})
            \gets
            \Call{Select}{i,\Omega_{i,t},\widehat{\Pi}_{i,t}^{(r-1)}}$.
            \State Set
            $\mu_{i,i,t}^{(r)}\gets j_{i,t}^{(r)}$ and
            $\beta_{i,i,t}^{(r)}\gets b_{i,t}^{(r)}$.
        \ElsIf{$\phi_{i,i,t}^{(r-1)}=1$}
            \State Preserve the fixed self-record:
      $ \mu_{i,i,t}^{(r)}\gets\mu_{i,i,t}^{(r-1)}, \
            \beta_{i,i,t}^{(r)}\gets\beta_{i,i,t}^{(r-1)}.$
        \Else
            \State Set
            $\mu_{i,i,t}^{(r)}\gets 0$ and
            $\beta_{i,i,t}^{(r)}\gets 0$.
        \EndIf
    \EndFor
   \State Each agent $i$ exchanges its pre-resolution self-record
$\bigl(\mu_{i,i,t}^{(r)},\beta_{i,i,t}^{(r)},\phi_{i,i,t}^{(r-1)}\bigr)$
with neighbors in $\mathcal{G}_t$.
    \ForAll{$i\in\mathcal{V}_t$}
       \State Update locally stored neighbor records:
$(\mu_{i,k,t}^{(r)},\beta_{i,k,t}^{(r)},\phi_{i,k,t}^{(r)})
\leftarrow
(\mu_{k,k,t}^{(r)},\beta_{k,k,t}^{(r)},\phi_{k,k,t}^{(r-1)})$,
for all $k\in\mathcal{N}_{i,t}$.
        \State
     $(\mu_{i,t}^{(r)},\beta_{i,t}^{(r)},\phi_{i,t}^{(r)})
        \gets$
        \Statex $\Call{ResolveAndUpdate}
{i,\mu_{i,t}^{(r)},\beta_{i,t}^{(r)},\phi_{i,t}^{(r)}}$.
    \EndFor

\EndFor

\State Output
$\Pi_t^{\rm D}
=
\{(i,\mu_{i,i,t}^{(R_t^{\max})})
\mid
\phi_{i,i,t}^{(R_t^{\max})}=1,\
\mu_{i,i,t}^{(R_t^{\max})}\neq0\}$.
\end{algorithmic}
\end{algorithm}

\begin{algorithm}[t]
\caption{$\textsc{Select}(i,\Omega_{i,t},\widehat{\Pi})$}
\label{alg:dagba_nominate}
\begin{algorithmic}[1]
\ForAll{$(i,j)\in\Omega_{i,t}$}
    \State Compute
    $b_{ij,t}=\Delta_t\bigl((i,j)\mid \widehat{\Pi}\bigr)$.
\EndFor
\State Select
$j_{i,t}\in
\arg\max_{j:(i,j)\in\Omega_{i,t}} b_{ij,t}$,
with ties broken by the ordered task index.
\State Set $b_{i,t}=b_{ij_{i,t},t}$.
\State \Return $(j_{i,t},b_{i,t})$.
\end{algorithmic}
\end{algorithm}

\begin{algorithm}[t]
\caption{$\textsc{ResolveAndUpdate}(i,\mu_i,\beta_i,\phi_i)$}
\label{alg:dagba_resolve}
\begin{algorithmic}[1]
\If{$\phi_{i,i}=1$ or $\mu_{i,i}=0$}
\State \Return $(\mu_i,\beta_i,\phi_i)$.
\EndIf
\State Construct the locally visible competitor set
$
\mathcal{C}_{i}
=
\{k\in\mathcal{N}_{i,t}^{+}
\mid
\phi_{i,k}=0,\ 
\mu_{i,k}=\mu_{i,i}\}.
$
\If{$\mathcal{C}_{i}=\varnothing$}
\State \Return $(\mu_i,\beta_i,\phi_i)$.
\EndIf
\State Let
$k^\star\in
\arg\max_{k\in\mathcal{C}_{i}}(\beta_{i,k},-k)$.
\If{$k^\star=i$}
\State Confirm the selected pair:
$\phi_{i,i}\gets 1$.
\Else
\State Clear the losing self-record:
$\mu_{i,i}\gets 0$, $\beta_{i,i}\gets 0$, and $\phi_{i,i}\gets 0$.
\EndIf
\State \Return $(\mu_i,\beta_i,\phi_i)$.
\end{algorithmic}
\end{algorithm}

Section~\ref{sec.2} formulates the per-epoch allocation as monotone submodular maximization over the energy-feasible ground set $\Omega_t$ under the partition-matroid constraint $\mathcal{S}_t$. Solving this problem in a distributed manner is challenging for three reasons. First, the admissible candidate space is time-varying because agents, tasks, residual energies, and motion-feasibility conditions all change across epochs. Second, due to submodularity, the marginal contribution of an agent-task pair depends on which agents have already been assigned to the same task, so bids cannot be treated as fixed edge weights ~\cite{krause2014SubmodularFM,qu2019distributed}. Third, under sparse and time-varying communication, no agent can maintain a globally synchronized view of the accepted assignment~\cite{gharesifard2017distributed,grimsman2018impact}.

To address these difficulties, we propose the \textbf{Distributed Adaptive Greedy Bidding Algorithm (DAGBA)}, a distributed implementation of marginal-greedy submodular maximization under local information constraints. At each inner round, agents locally evaluate conditional marginal gains, exchange only locally visible bidding information, resolve conflicts among agents that preselect the same task, and update their local views of fixed assignments. Thus, the bidding mechanism is the communication-efficient realization of the greedy marginal-selection rule~\cite{krause2014SubmodularFM,qu2019distributed}.

DAGBA maintains local information states rather than a shared global assignment state. At inner round $r$, each agent $i$ keeps three local arrays $\mu_{i,t}^{(r)}=\{\mu_{i,k,t}^{(r)}\}_{k\in\mathcal{V}_t},\
\beta_{i,t}^{(r)}=\{\beta_{i,k,t}^{(r)}\}_{k\in\mathcal{V}_t},\
\phi_{i,t}^{(r)}=\{\phi_{i,k,t}^{(r)}\}_{k\in\mathcal{V}_t}$.
The entry $\mu_{i,k,t}^{(r)}$ represents agent $i$'s local view of the task selected by agent $k$, $\beta_{i,k,t}^{(r)}$ is the associated bid value, and $\phi_{i,k,t}^{(r)}\in\{0,1\}$ indicates whether agent $k$'s assignment is regarded as fixed in agent $i$'s local view. The self-state $(\mu_{i,i,t}^{(r)},\beta_{i,i,t}^{(r)},\phi_{i,i,t}^{(r)})$ represents agent $i$'s own selected task, bid value, and fixed status. Neighbor entries are refreshed only through one-hop communication over the current graph $\mathcal{G}_t$.

For marginal-gain evaluation, each agent converts its locally fixed information state into the set-valued estimate
$\widehat{\Pi}_{i,t}^{(r)} = \{(k,\mu_{i,k,t}^{(r)})
\mid \phi_{i,k,t}^{(r)}=1, \mu_{i,k,t}^{(r)}\neq 0\}$.
This estimate is the assignment information locally available to agent $i$ when computing conditional marginal gains, not a globally shared assignment. Under sparse communication, different agents may hold different estimates. This local-estimate structure avoids the need for global consensus over a time-varying communication graph.

For any admissible pair $\pi=(i,j)\in\Omega_t$ and any assignment set $\Pi\subseteq\Omega_t$, define the conditional marginal utility $ \Delta_t(\pi\mid \Pi) = U_t(\Pi\cup\{\pi\})-U_t(\Pi)$.
Since $U_t$ is submodular, the marginal contribution of assigning agent $i$ to task $j$ depends on the agents already assigned to task $j$ in the local estimate~\cite{krause2014SubmodularFM,qu2019distributed}. Therefore, DAGBA recomputes bids against the latest locally fixed assignment estimate rather than using static edge weights.

For each agent $i\in\mathcal{V}_t$, define its local admissible subset
\begin{equation}
\Omega_{i,t}
=
\{(i,j)\in\Omega_t\mid j\in\mathcal{M}_t\}.
    \label{eq:agent_subset}
\end{equation}
Thus, $\Omega_{i,t}$ contains exactly the task choices currently executable by agent $i$. Since $\Omega_t$ already incorporates the residual-energy and motion-feasibility conditions, infeasible trajectories and energy-infeasible pairs are excluded before bidding.

At inner round $r$, if agent $i$ is not fixed in its own local view, i.e., $\phi_{i,i,t}^{(r-1)}=0$, and $\Omega_{i,t}\neq\emptyset$, it constructs $\widehat{\Pi}_{i,t}^{(r-1)}$ and selects a task by
$j_{i,t}^{(r)} \in \arg\max_{j:(i,j)\in\Omega_{i,t}} \Delta_t\big((i,j)\mid \widehat{\Pi}_{i,t}^{(r-1)}\big)$, and
 $b_{i,t}^{(r)} = \Delta_t\big((i,j_{i,t}^{(r)})\mid \widehat{\Pi}_{i,t}^{(r-1)}\big)$,
where ties are broken by the ordered task index. The self-state is then updated as $\mu_{i,i,t}^{(r)}=j_{i,t}^{(r)}$ and $\beta_{i,i,t}^{(r)}=b_{i,t}^{(r)}$. If agent $i$ has already been fixed, its self-state is preserved and it no longer submits a new bid. If no admissible task is available, it submits a null state.

After the bidding step, each agent exchanges only its self-state with its current neighbors. Using the refreshed neighbor entries, agent $i$ resolves conflicts among locally visible agents that select the same task. Specifically, if agent $i$ is not fixed and has selected a nonzero task, it forms the locally visible competitor set $\mathcal{C}_{i,t}^{(r)} = \{k\in\mathcal{N}_{i,t}^{+}
\mid \phi_{i,k,t}^{(r-1)}=0,\ \mu_{i,k,t}^{(r)}=\mu_{i,i,t}^{(r)}\}$,
where $\mathcal{N}_{i,t}^{+}=\mathcal{N}_{i,t}\cup\{i\}$ is the closed neighborhood of agent $i$. 
The locally visible agent with the lexicographically largest pair
$(\beta_{i,k,t}^{(r)},-k)$ is regarded as the local winner. If the local
winner is agent $i$ itself, agent $i$ confirms its selected pair by setting
$\phi_{i,i,t}^{(r)}=1$. Otherwise, agent $i$ clears its tentative self-record
and remains unfixed.

The complete procedure is summarized in Algorithms~\ref{alg:dagba_main}-\ref{alg:dagba_resolve}. Algorithm~\ref{alg:dagba_main} describes the epoch-level coordination loop. Algorithm~\ref{alg:dagba_nominate} implements the conditional marginal-gain selection step, and Algorithm~\ref{alg:dagba_resolve} implements the local conflict-resolution and information-state update step.
The assignment returned by DAGBA at epoch $t$ is induced by the fixed self-record at the end of the prescribed inner-round budget:
$\Pi_t^{\rm D}
=
\left\{
(i,\mu_{i,i,t}^{(R_t^{\max})})
 \middle| 
\phi_{i,i,t}^{(R_t^{\max})}=1,\
\mu_{i,i,t}^{(R_t^{\max})}\neq0
\right\}$. Equivalently, each fixed agent executes its fixed pair
$(i,\mu_{i,i,t}^{(R_t^{\max})})$. Since each agent fixes at most one self-state, the output satisfies the per-agent partition constraint. Multiple agents may still be assigned to the same task, either across different rounds or within the same round when their mutual conflict is not visible under the current sparse communication graph.

The design of DAGBA reflects the structure of \cref{prob:stage_allocation}. The local admissible subset $\Omega_{i,t}$ handles the time-varying candidate space induced by active-set variation, motion feasibility, and residual-energy constraints. Conditional marginal bidding implements the submodular greedy principle by evaluating each candidate relative to the latest locally fixed assignment estimate. Neighbor-based conflict resolution coordinates locally competing agents without requiring global consensus. The agent-wise local information states allow the allocation process to proceed even when the communication graph is sparse or disconnected.

\section{Performance Analysis of DAGBA}
\label{sec.4}

This section analyzes the per-epoch performance of DAGBA. At a fixed epoch $t$, the tuple $(\mathcal{V}_t,\mathcal{M}_t,\mathcal{G}_t,\Omega_t,\mathcal{S}_t,U_t)$ is determined by the current system information. Let $\Pi_t^\star \in \arg\max_{\Pi\in\mathcal{S}_t} U_t(\Pi)$ denote an optimal assignment for Problem~\ref{prob:stage_allocation} at epoch $t$, and let $\Pi_t^{\rm D}$ denote the assignment returned by DAGBA. The guarantees compare the DAGBA assignment $\Pi_t^{\rm D}$ with an optimal solution $\Pi_t^\star$ of the same-epoch Problem~\ref{prob:stage_allocation}. Across epochs, the executed assignment influences future stage problems through the residual-energy update~\eqref{eq:energy_update}, agent motion, task completion, and the resulting communication topology.

\subsection{Analysis Setup and Feasibility}
\label{subsec:analysis_setup_feasibility}

By~\eqref{eq:agent_subset}, each active agent $i\in\mathcal{V}_t$ has a local admissible subset $\Omega_{i,t}=\{(i,j)\in\Omega_t\mid j\in\mathcal{M}_t\}$. Let $r_t=|\{i\in\mathcal{V}_t\mid \Omega_{i,t}\neq\emptyset\}|$ denote the number of active agents that have at least one admissible candidate pair at epoch $t$. {Since the sets $\Omega_{i,t}$ are pairwise disjoint and $\mathcal{S}_t$ selects at most one element from each $\Omega_{i,t}$, the set system $(\Omega_t,\mathcal{S}_t)$ is a partition matroid by Definition~\ref{def:partition_matroid}, induced by the partition $\{\Omega_{i,t}\}_{i\in\mathcal{V}_t:\Omega_{i,t}\neq\emptyset}$.}

Define the set of agent-task pairs newly fixed at inner round $r$ by
\begin{equation}
\mathcal{A}_t^{(r)}
=
\left\{
(i,\mu_{i,i,t}^{(r)})
\ \middle|\
\phi_{i,i,t}^{(r)}=1,\
\phi_{i,i,t}^{(r-1)}=0,\
\mu_{i,i,t}^{(r)}\neq 0
\right\}.
\label{eq:newly_fixed_pairs}
\end{equation}
Then the assignment returned by DAGBA is
\begin{equation}
{
\Pi_t^{\rm D}
=
\bigcup_{r=1}^{R_t^{\max}}\mathcal{A}_t^{(r)}
=
\left\{
(i,\mu_{i,i,t}^{(R_t^{\max})})
 \middle| 
\phi_{i,i,t}^{(R_t^{\max})}=1,\
\mu_{i,i,t}^{(R_t^{\max})}\neq0
\right\}.}
\label{eq:dagba_output_analysis}
\end{equation}
{If $R_t^{\max}\ge r_t$, all active agents with nonempty admissible subsets become fixed within at most $r_t$ inner rounds, while the remaining inner rounds only preserve the fixed self-records.}

\begin{proposition}[Feasibility and finite-round convergence]
\label{prop:feasibility_termination}
{For every epoch $t$, DAGBA returns an assignment
$\Pi_t^{\rm D}\in\mathcal{S}_t$ after $R_t^{\max}$ inner rounds, and the residual-energy update satisfies
$E_{i,t+1}\ge0$ for all $i\in\mathcal{V}_t$. If
$R_t^{\max}\ge r_t$, then all active agents with nonempty local admissible subsets become fixed within at most $r_t$ inner rounds; all subsequent inner rounds preserve their fixed self-records.}
\end{proposition}
\begin{proof}
{\emph{i) Feasibility.} By construction, every pair in $\Pi_t^{\rm D}$ belongs to $\Omega_t$. Moreover, by the definition~\eqref{eq:dagba_output_analysis}, $\Pi_t^{\rm D}$ contains at most one pair $(i,j)$ for each agent $i\in\mathcal{V}_t$, so $|\Pi_t^{\rm D}\cap\Omega_{i,t}|\le 1$ for all $i\in\mathcal{V}_t$. Hence $\Pi_t^{\rm D}\in\mathcal{S}_t$.}

\emph{ii) Energy safety.} If agent $i$ is unassigned, then $E_{i,t+1}=E_{i,t}\ge0$. If agent $i$ is assigned to task $j$, then $(i,j)\in\Omega_t$ implies $\widehat{C}_{ij,t}\le E_{i,t}$ by~\eqref{eq:agent-task}. By Assumption~\ref{ass:cost_upper_bound}, $e_{ij,t}\le\widehat{C}_{ij,t}$, and since $\Pi_t^{\rm D}\in\mathcal{S}_t$ assigns agent $i$ to at most one task, $E_{i,t+1}\ge E_{i,t}-\widehat{C}_{ij,t}\ge 0$.

\emph{iii) Finite-round convergence.} The inner loop contains $R_t^{\max}$ rounds by construction. When $R_t^{\max}\ge r_t$, we show that at least one previously unconfirmed agent confirms in each round. Among all unfixed agents that select the same task $j$, consider the subgraph of $\mathcal{G}_t$ they induce. In every connected component of this subgraph, the agent with the lexicographically largest pair $(\beta_{k,k,t}^{(r)},-k)$ sees no same-task neighbor with a larger value after the neighbor exchange, and hence confirms by Algorithm~\ref{alg:dagba_resolve}. {Therefore, at least one previously unfixed agent becomes fixed in every such round, so all agents with nonempty admissible subsets are fixed within at most $r_t$ inner rounds and remain fixed thereafter.}
\end{proof}

\subsection{Sequential Greedy Approximation}
\label{subsec:sequential_greedy}

{This subsection identifies a regime in which DAGBA reduces to the classical sequential greedy algorithm and therefore inherits its approximation guarantee.}
For any feasible assignment $\Pi\in\mathcal{S}_t$, define the feasible augmentation set
$\mathcal{B}_t(\Pi) = \left\{ \pi\in\Omega_t\setminus\Pi
\mid \Pi\cup\{\pi\}\in\mathcal{S}_t \right\}$.
A centralized sequential greedy algorithm starts from $G_{0,t}=\emptyset$ and, at step $k$, selects
\begin{equation}
\begin{aligned}
g_{k,t}
&\in
\arg\max_{\pi\in\mathcal{B}_t(G_{k-1,t})}
\Delta_t(\pi\mid G_{k-1,t}),\\
G_{k,t}
&=G_{k-1,t}\cup\{g_{k,t}\}.
\end{aligned}
\label{eq:centralized_greedy_rule}
\end{equation}
The process stops when $\mathcal{B}_t(G_{k-1,t})=\emptyset$. {Since $(\Omega_t,\mathcal{S}_t)$ is a partition matroid, this is the standard greedy procedure for monotone submodular maximization under a matroid constraint~\cite{fisher1978analysis}.}

{DAGBA reproduces the centralized sequential greedy order provided that, before each successful fixing step, (i)~all unfixed agents share the same fixed-pair set $G\subseteq\Omega_t$, (ii)~each unfixed agent evaluates its candidate marginal gains with respect to $G$, and (iii)~exactly one pair
$\pi^\star\in\arg\max_{\pi\in\mathcal{B}_t(G)}\Delta_t(\pi\mid G)$
is fixed and added to $G$. These conditions characterize the sequential-greedy setting considered in this subsection, in which all unfixed agents evaluate their bids with respect to the same fixed-pair set and exactly one pair attaining the globally largest current marginal gain is fixed at each step.}

\begin{lemma}
\label{lem:dagba_sequential_reduction}
{If Conditions~(i)-(iii) above hold}, the output of DAGBA at epoch $t$ can be ordered as $\Pi_t^{\rm D}=\{g_{1,t},\ldots,g_{K,t}\}$ such that the ordering satisfies the centralized greedy rule in~\eqref{eq:centralized_greedy_rule}.
\end{lemma}
\begin{proof}
Let $G_{k-1,t}=\{g_{1,t},\ldots,g_{k-1,t}\}$ be the set of pairs fixed before the $k$-th successful self-fixing event. {By condition (i), each unfixed agent evaluates its marginal bid with respect to $G_{k-1,t}$.} For each unfixed agent, the local admissible candidates are exactly the feasible augmentations associated with that agent's partition class. Since each agent submits a bid for its best candidate under $G_{k-1,t}$, the largest submitted bid coincides with the largest marginal gain over the full feasible augmentation set $\mathcal{B}_t(G_{k-1,t})$. {By condition (iii), the unique fixed pair at this step is therefore an element of
$\arg\max_{\pi\in\mathcal{B}_t(G_{k-1,t})}
\Delta_t(\pi\mid G_{k-1,t})$,
which is exactly the centralized greedy rule in~\eqref{eq:centralized_greedy_rule}.} Repeating this argument for all successful self-fixing events gives the desired ordering.
\end{proof}

\begin{theorem}[Sequential greedy approximation]
\label{thm:sequential_greedy}
{If Conditions~(i)-(iii) above hold},
\begin{equation}
U_t(\Pi_t^{\rm D})
\ge
\frac{1}{2}U_t(\Pi_t^\star).
\label{eq:sequential_half_bound}
\end{equation}
Moreover, if the total curvature of $U_t$ is $c_t\in[0,1]$, then
\begin{equation}
U_t(\Pi_t^{\rm D})
\ge
\frac{1}{1+c_t}U_t(\Pi_t^\star).
\label{eq:sequential_curvature_bound}
\end{equation}
\end{theorem}
\begin{proof}
{By Lemma~\ref{lem:dagba_sequential_reduction}, the output of DAGBA coincides with the output of the centralized sequential greedy algorithm applied to Problem~\ref{prob:stage_allocation}.} Since $U_t$ is normalized, monotone nondecreasing, and submodular, the classical greedy guarantee for monotone submodular maximization under a matroid constraint gives~\eqref{eq:sequential_half_bound}~\cite{fisher1978analysis}. If the total curvature of $U_t$ is $c_t$, the curvature-refined greedy guarantee gives~\eqref{eq:sequential_curvature_bound}~\cite{conforti1984submodular}.
\end{proof}

{Theorem~\ref{thm:sequential_greedy} shows that DAGBA recovers the standard greedy approximation guarantee when the distributed execution has complete visibility of previously fixed pairs and accepts one globally best marginal bid at a time. In sparse communication graphs, these conditions generally do not hold, and the next subsection quantifies the resulting performance loss.}


\subsection{Approximation under Local Information}
\label{subsec:local_information_approximation}

{In the standard execution of DAGBA, each agent computes marginal bids using its local assignment estimate, and several pairs may be fixed in the same inner round. This subsection derives the corresponding approximation guarantee.}
We first state a consistency condition on the local assignment estimates used in the analysis.
\begin{lemma}[Confirmed local estimates]
\label{lem:confirmed_local_estimates}
Under Algorithm~\ref{alg:dagba_main} with the confirmed update rule in Algorithm~\ref{alg:dagba_resolve}, every local assignment estimate used in marginal-gain evaluation contains only pairs that belong to the final DAGBA assignment. That is, for every active agent $i$ and every inner round $r$,
$\widehat{\Pi}_{i,t}^{(r-1)}\subseteq \Pi_t^{\rm D}$.
\end{lemma}
\begin{proof}
Consider any pair $(k,\mu_{i,k,t}^{(r-1)})\in \widehat{\Pi}_{i,t}^{(r-1)}$. By the definition of the local assignment estimate, $\phi_{i,k,t}^{(r-1)}=1$ and $\mu_{i,k,t}^{(r-1)}\neq 0$.
Under Algorithm~\ref{alg:dagba_resolve}, an agent can set only its own fixed indicator from zero to one. Hence $\phi_{i,k,t}^{(r-1)}=1$ must have originated from a confirmed fixed status of agent $k$ that was propagated to agent $i$ through the exchanged self-records. Once agent $k$ confirms its assignment, Algorithm~\ref{alg:dagba_main} preserves its self-record in later inner rounds. Therefore, at the final round $R_t^{\max}$, agent $k$ remains fixed with the same selected task, and $(k,\mu_{i,k,t}^{(r-1)})\in \Pi_t^{\rm D}$.
\end{proof}

For each pair $\pi=(i,j)\in\mathcal{A}_t^{(r)}$ newly fixed at inner round $r$, define the local bid value $b_t(\pi) =
\beta_{i,i,t}^{(r)} = \Delta_t \left(
\pi\mid\widehat{\Pi}_{i,t}^{(r-1)}
\right)$,
{and the corresponding local marginal sum $L_t = \sum_{r=1}^{R_t^{\max}}
\sum_{\pi\in\mathcal{A}_t^{(r)}}b_t(\pi)$.}
If $U_t(\Pi_t^{\rm D})>0$, define the local-information factor
\begin{equation}
\eta_t
=
\frac{L_t}{U_t(\Pi_t^{\rm D})}.
\label{eq:local_information_factor}
\end{equation}
The quantity $\eta_t$ measures how much the sum of locally evaluated marginal bids exceeds the final utility obtained by DAGBA. Under full sequential information, the local marginal sum telescopes and $\eta_t=1$. Under local information, missing fixed pairs may cause agents to overestimate marginal gains, and {$\eta_t$ quantifies this overestimation.}

\begin{theorem}[Approximation under local information]
\label{thm:local_information_approximation}
Suppose that Algorithm~\ref{alg:dagba_main} is executed with the confirmed update rule in Algorithm~\ref{alg:dagba_resolve} and a round budget satisfying $R_t^{\max}\ge r_t$. If $U_t(\Pi_t^{\rm D})>0$, then
\begin{equation}
U_t(\Pi_t^{\rm D})
\ge
\frac{1}{1+\eta_t}
U_t(\Pi_t^\star).
\label{eq:local_information_bound}
\end{equation}
If $U_t(\Pi_t^{\rm D})=0$, then $U_t(\Pi_t^\star)=0$.
\end{theorem}
\begin{proof}
{Since $U_t$ is monotone submodular, the following standard inequality holds for any $A,B\subseteq\Omega_t$ \cite{nemhauser1978analysis}. Indeed, order $A\setminus B=\{\pi_1,\ldots,\pi_m\}$ and let $B_\ell=B\cup\{\pi_1,\ldots,\pi_\ell\}$. Then
\begin{align*}
U_t(A)
&\le
U_t(A\cup B)
\ \text{(monotonicity)}\\
&=
U_t(B)+\sum_{\ell=1}^{m}\Delta_t(\pi_\ell\mid B_{\ell-1})
\ \text{(telescoping)}\\
&\le
U_t(B)+\sum_{\ell=1}^{m}\Delta_t(\pi_\ell\mid B)
\ \text{(submodularity, } B\subseteq B_{\ell-1}\text{)}\\
& = U_t(B)+\sum_{\pi\in A\setminus B}\Delta_t(\pi\mid B).
\end{align*}
Applying this inequality with $A=\Pi_t^\star$ and $B=\Pi_t^{\rm D}$ gives $U_t(\Pi_t^\star)\le U_t(\Pi_t^{\rm D})+\sum_{\pi\in\Pi_t^\star\setminus\Pi_t^{\rm D}}\Delta_t(\pi\mid \Pi_t^{\rm D})$.}

Consider any pair $\pi^\star=(i,j^\star)\in\Pi_t^\star\setminus\Pi_t^{\rm D}$. Since $\pi^\star\in\Omega_{i,t}$ and DAGBA fixes every active agent with a nonempty local admissible subset, agent $i$ is fixed by DAGBA to some pair $\pi_i^{\rm D}=(i,j_i^{\rm D})\in\Pi_t^{\rm D}$ at inner round $r_i$. By Lemma~\ref{lem:confirmed_local_estimates}, $\widehat{\Pi}_{i,t}^{(r_i-1)}\subseteq\Pi_t^{\rm D}$. {By submodularity and the greedy selection rule,}
$\Delta_t(\pi^\star\mid \Pi_t^{\rm D})
\le
\Delta_t(\pi^\star\mid \widehat{\Pi}_{i,t}^{(r_i-1)})
\le
\Delta_t(\pi_i^{\rm D}\mid \widehat{\Pi}_{i,t}^{(r_i-1)})
=
b_t(\pi_i^{\rm D})$,
{where the second inequality uses $\pi^\star\in\Omega_{i,t}$ and the fact that agent $i$ maximized the conditional marginal gain over $\Omega_{i,t}$.}

{Since both $\Pi_t^\star$ and $\Pi_t^{\rm D}$ satisfy the partition-matroid constraint, each agent contributes at most one pair to each assignment, so each selected pair in $\Pi_t^{\rm D}$ is charged at most once. Therefore $\sum_{\pi\in\Pi_t^\star\setminus\Pi_t^{\rm D}}\Delta_t(\pi\mid \Pi_t^{\rm D})\le L_t$, which gives $U_t(\Pi_t^\star)\le U_t(\Pi_t^{\rm D})+L_t$.}
If $U_t(\Pi_t^{\rm D})>0$, substituting the definition of $\eta_t$ in~\eqref{eq:local_information_factor} proves~\eqref{eq:local_information_bound}. If $U_t(\Pi_t^{\rm D})=0$, then all selected local bids are zero, and the same charging argument gives $U_t(\Pi_t^\star)\le0$. Since $U_t$ is nonnegative, $U_t(\Pi_t^\star)=0$.
\end{proof}

{Theorem~\ref{thm:local_information_approximation} separates the effect of local information from the feasibility structure of the problem. The factor $\eta_t$ equals one in the full-information sequential case and can be larger when agents compute bids without observing some previously fixed pairs. Thus, performance degradation is caused by marginal-gain overestimation under local assignment estimates, not by violation of the partition-matroid constraint. The following corollary gives a conservative bound in terms of $K_t$, $R_t^{\max}$, and $s_{\max,t}$.}

\begin{corollary}[Computable bound]
\label{cor:computable_bound}
{
Let $K_t=|\Pi_t^{\rm D}|$ and
$s_{\max,t} = \max_{1\le r\le R_t^{\max}}
|\mathcal{A}_t^{(r)}|$.}
Under the conditions of Theorem~\ref{thm:local_information_approximation},
\begin{equation}
U_t(\Pi_t^{\rm D})
\ge
\frac{1}{K_t+1}U_t(\Pi_t^\star)
\ge
{
\frac{1}{s_{\max,t}R_t^{\max}+1}U_t(\Pi_t^\star).}
\label{eq:computable_bound}
\end{equation}
\end{corollary}
\begin{proof}
{For any fixed pair $\pi\in\Pi_t^{\rm D}$, by submodularity and monotonicity, $b_t(\pi)=\Delta_t(\pi\mid\widehat{\Pi})\le\Delta_t(\pi\mid\emptyset)=U_t(\{\pi\})\le U_t(\Pi_t^{\rm D})$. Therefore $L_t\le K_t U_t(\Pi_t^{\rm D})$, which implies $\eta_t\le K_t$. Substituting into~\eqref{eq:local_information_bound} yields the first inequality. Since
$K_t\le s_{\max,t}R_t^{\max}$,
the second inequality follows.}
\end{proof}

{The bound in Corollary~\ref{cor:computable_bound} is conservative but directly computable from the prescribed round budget and the observed number of pairs fixed in each inner round.}

\subsection{Computational Complexity}
\label{subsec:computational_complexity}

{We analyze the computational and communication costs of DAGBA at a fixed epoch $t$.} Let $N_t=|\mathcal{V}_t|$, $M_t=|\mathcal{M}_t|$, and $m_{i,t}=|\Omega_{i,t}|$ for each active agent $i\in\mathcal{V}_t$. Then $|\Omega_t|=\sum_{i\in\mathcal{V}_t}m_{i,t}$. Let $C_{\rm plan}$ denote the cost of evaluating the motion feasibility and predicted execution cost for one candidate agent-task pair, and let $C_\Delta$ denote the cost of evaluating one conditional marginal utility $\Delta_t((i,j)\mid\widehat{\Pi})$.

{The admissible ground set is constructed before the inner bidding process starts. If each active agent checks all active tasks, the construction cost is $O(N_tM_t C_{\rm plan})$. At each inner round, every unfixed agent evaluates the conditional marginal utility of its admissible pairs, giving $O(|\Omega_t|C_\Delta)$ marginal-evaluation cost. The local conflict-resolution step requires $O(N_t+|\mathcal{E}_t|)$ operations per round. In each inner round, every active agent sends its self-record to its neighbors, giving $O(|\mathcal{E}_t|)$ messages per round.}

\begin{proposition}[Complexity of DAGBA]
\label{prop:complexity_dagba}
{At epoch $t$, DAGBA has worst-case computational complexity
$O \left( N_tM_tC_{\rm plan} + R_t^{\max}
\left[ |\Omega_t|C_\Delta+N_t+|\mathcal{E}_t|
\right] \right)$. Since $|\Omega_t|\le N_tM_t$, a coarse upper bound is
$O \left( N_tM_tC_{\rm plan} + R_t^{\max} \left[ N_tM_tC_\Delta+N_t+|\mathcal{E}_t|
\right] \right)$. 
The worst-case communication complexity at epoch $t$ is
$O(R_t^{\max}|\mathcal{E}_t|)$ messages.}
\end{proposition}
\begin{proof}
{The construction of $\Omega_t$ requires evaluating motion feasibility and predicted execution cost for candidate pairs, giving $O(N_tM_t C_{\rm plan})$. During each inner round, each agent $i$ evaluates $\Delta_t((i,j)\mid\widehat{\Pi}_{i,t}^{(r-1)})$ for all $(i,j)\in\Omega_{i,t}$, giving $O(|\Omega_t|C_\Delta)$ total marginal-evaluation cost. The neighbor exchange and conflict-resolution step requires $O(N_t+|\mathcal{E}_t|)$ local operations per round. Multiplying the per-round costs by $R_t^{\max}$ and adding the admissible-set construction cost yields the stated worst-case computational complexity. Since each edge carries a constant-size message per inner round, the worst-case communication complexity is
$O(R_t^{\max}|\mathcal{E}_t|)$.}
\end{proof}

{Under Algorithm~\ref{alg:dagba_main}, the local storage required by agent $i$ is $O(N_t+m_{i,t})$. The complexity depends on the current system size through $N_t$, $M_t$, $|\Omega_t|$, and $|\mathcal{E}_t|$. Agents and tasks inactive at epoch $t$ do not enter the computation.}


\section{Applications and Simulations}\label{sec.5}

This section evaluates DAGBA in a cooperative observation and tracking mission. We assess stage allocation quality, finite-horizon task-completion performance, communication sensitivity, runtime, and the role of the main algorithmic components under dynamic agent participation and limited communication. All simulations are implemented in MATLAB R2022b.

\begin{figure}[t]
    \centering
    \begin{subfigure}{\linewidth}
        \centering
        \includegraphics[width=\linewidth]{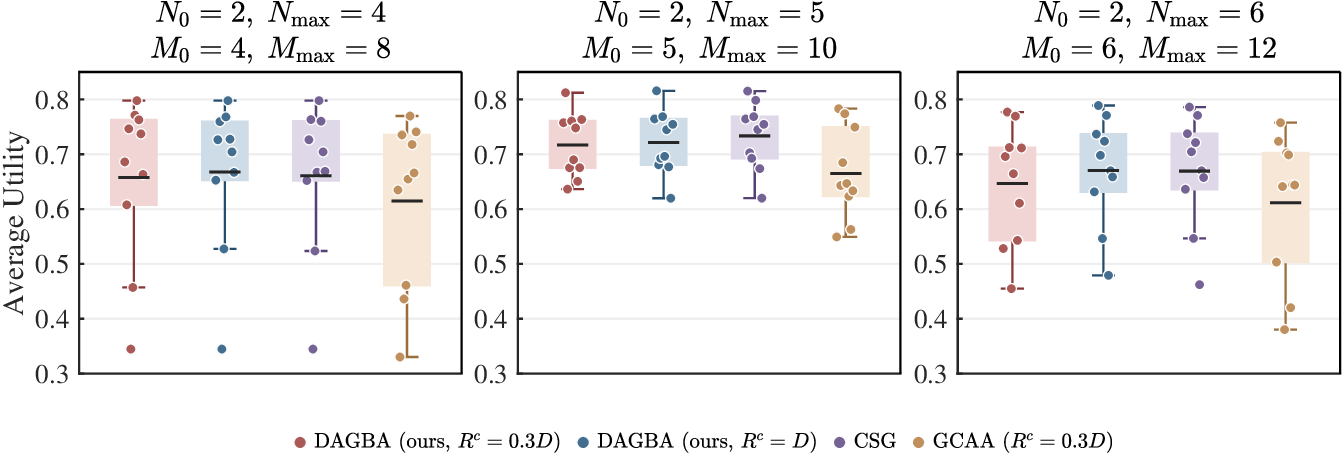}
        \caption{Average utility.}
        \label{fig:mission_average_utility}
    \end{subfigure}
    \vspace{0.25em}
    \begin{subfigure}{\linewidth}
        \centering
        \includegraphics[width=\linewidth]{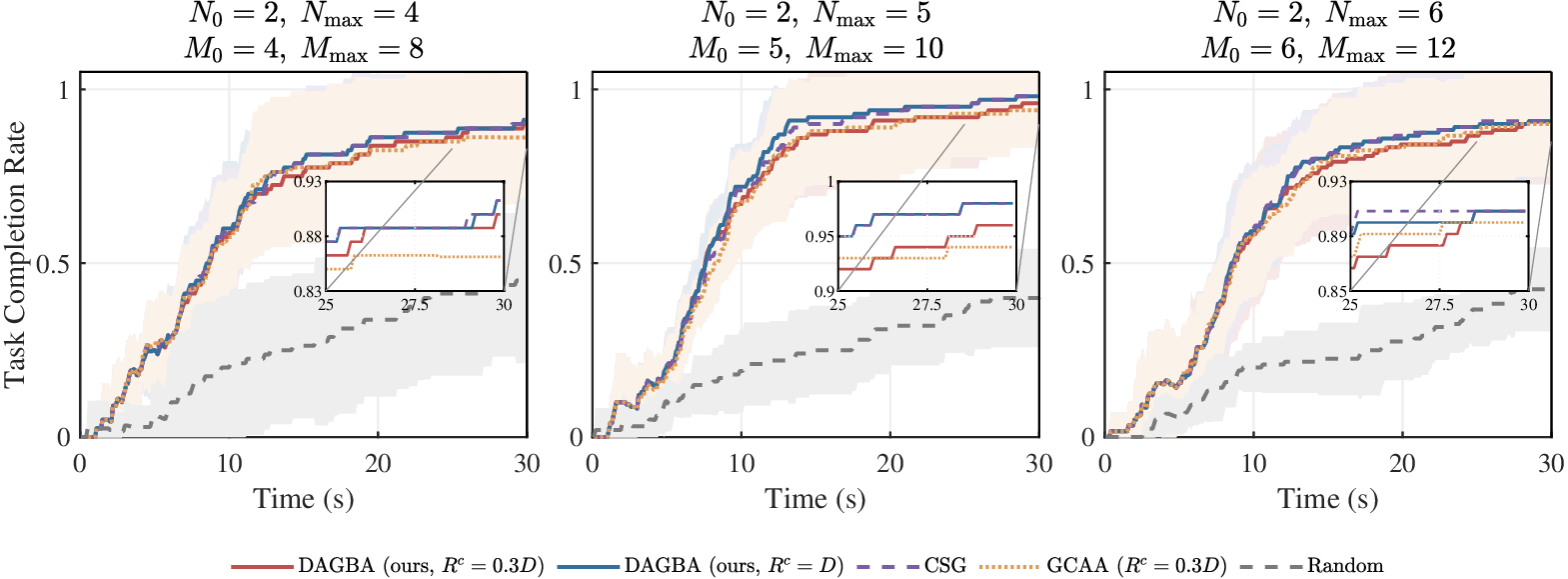}
        \caption{Task completion rate.}
        \label{fig:mission_completion}
    \end{subfigure} 
    \vspace{0.25em}
    \begin{subfigure}{\linewidth}
        \centering
        \includegraphics[width=\linewidth]{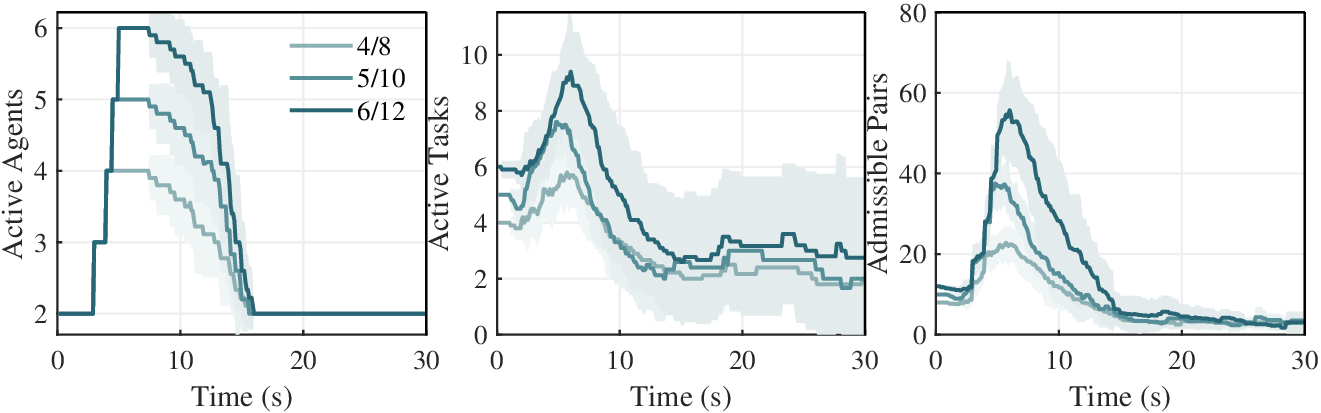}
        \caption{{System profile.}}
        \label{fig:mission_open_profile}
    \end{subfigure}
\caption{{Mission performance and system characteristics under three problem scales. (a) reports the distribution of the workload-normalized average utility over $10$ independent trials. (b) reports the task completion rate over time, where the curves represent the sample mean and the shaded regions denote the standard error. (c) shows the evolution of active agents, active tasks, and admissible pairs. DAGBA under sparse communication maintains utility and completion progress close to the complete-communication and centralized greedy references while operating under time-varying participants and feasible candidate pairs.}}
\label{fig:revised_main_performance}
\end{figure}

\subsection{Simulation Setup}
\label{subsec:simulation_setup}

We consider a cooperative observation and tracking mission {deployed over a bounded two-dimensional workspace $\mathcal{W}=[-10,10]^2~\mathrm{m}$}. The simulation instantiates the {per-epoch} allocation model in Section~\ref{sec.2}: at each epoch $t$, the active agent set $\mathcal{V}_t$, active task set $\mathcal{M}_t$, communication graph $\mathcal{G}_t$, residual-energy states, admissible ground set $\Omega_t$, and feasible family $\mathcal{S}_t$ are updated before the allocation decision is made. Each active agent $i\in\mathcal{V}_t$ has state $x_{i,t}=(\mathbf{p}_{i,t},\mathbf{v}_{i,t})^\top\in\mathbb{R}^{4}$, where $\mathbf{p}_{i,t}\in\mathbb{R}^{2}$ and $\mathbf{v}_{i,t}\in\mathbb{R}^{2}$ denote its position and velocity, respectively. The control input satisfies $\Vert u_{i,t}\Vert\le u_{\max}$. Each active task $j\in\mathcal{M}_t$ corresponds to a moving target with state $y_{j,t}=(\mathbf{q}_{j,t},\mathbf{w}_{j,t})^\top\in\mathbb{R}^{4}$, where $\mathbf{q}_{j,t}\in\mathbb{R}^{2}$ is the target position. A task is completed and removed from $\mathcal{M}_t$ once at least one assigned agent enters its observation radius $R^{\mathrm{obs}}$. 

Each potential agent becomes active at a prescribed entry epoch and may later fail or depart from the system. Each potential task becomes active at a prescribed arrival epoch and is removed after completion. At epoch $t$, the active sets $\mathcal{V}_t$ and $\mathcal{M}_t$ are determined by this event schedule and the task-completion status. Throughout the mission horizon, the active sets satisfy $|\mathcal{V}_t|\le N_{\max}$ and $|\mathcal{M}_t|\le M_{\max}$. Agent and task arrivals introduce new candidate agent-task pairs, whereas agent failures and task completions remove previously available pairs from the current {admissible ground set}.

The communication graph $\mathcal{G}_t=(\mathcal{V}_t,\mathcal{E}_t)$ is generated from the current agent positions. Two active agents are connected if their inter-agent distance is no larger than the communication radius $R^c$. Let $D$ denote the diameter of $\mathcal{W}$. The complete-communication setting uses $R^c=D$, under which all active agents are mutually reachable. The sparse-communication setting uses $R^c=\rho D$ with $\rho<1$, so that agents exchange bidding and assignment-status information only with nearby neighbors.

For each candidate pair $(i,j)\in\mathcal{V}_t\times\mathcal{M}_t$, the lower-level planner computes the predicted execution cost $\widehat{C}_{ij,t}$ from the current agent state, target state, and local motion-feasibility conditions. If no feasible local trajectory exists, we set $\widehat{C}_{ij,t}=+\infty$. The admissible ground set and feasible family are then constructed according to~\eqref{eq:agent-task} and~\eqref{eq:feasible_family}, respectively. After the selected assignment is executed for one epoch, residual energies are updated according to~\eqref{eq:energy_update}. Energy consumed by the current assignment may reduce the admissible ground set at subsequent epochs, coupling consecutive per-epoch allocation problems through residual resources.

The stage utility is constructed from cooperative observation probabilities. For each active agent-task pair $(i,j)\in V_t\times M_t$, the per-epoch observation success probability is modeled as
 $\mathcal{P}_{ij,t}  = \exp\bigl(-\lambda_j\Vert \mathbf{p}_{i,t}-\mathbf{q}_{j,t}\Vert\bigr),  \ \lambda_j>0$ .
For each task $j\in\mathcal{M}_t$ and each agent subset $C\subseteq\mathcal{V}_t$, the task utility is 
\begin{equation}
F_{j,t}(C)
=
\omega_j
\left(
1-\prod_{i\in C}(1-\mathcal{P}_{ij,t})
\right),
\
\omega_j>0.
\label{eq:sim_task_utility}
\end{equation}
The allocation utility follows~\eqref{eq:stage_utility}. The union-of-success form in~\eqref{eq:sim_task_utility} is normalized, monotone nondecreasing, and submodular in $C$, {as each additional agent increases the observation probability with diminishing marginal improvement}~\cite{krause2014SubmodularFM}. Therefore, the simulated utility is consistent with the monotone submodular objective used in Problem~\ref{prob:stage_allocation}.

The common simulation parameters are as follows. The initial residual energy is $E_i^0=300$, the acceleration bound is $u_{\max}=1.2\,\mathrm{m/s^2}$, the observation radius is $R^{\mathrm{obs}}=0.5\,\mathrm{m}$, and $\lambda_j=0.1$ for all tasks. 
Agents re-plan at $10\,\mathrm{Hz}$, corresponding to the physical-time sampling interval $\Delta\tau=0.1\,\mathrm{s}$.
The terminal simulation time is
$t_f=30\,\mathrm{s}$, and hence each trial contains $T_{\mathrm{sim}} = \frac{t_f}{\Delta\tau} = 300$ decision epochs.
Each configuration is evaluated over $10$ independent Monte Carlo trials. Curves report the sample mean, and shaded regions or error bars denote the standard error.

\begin{figure}[t]
    \centering
    \includegraphics[width=\linewidth]{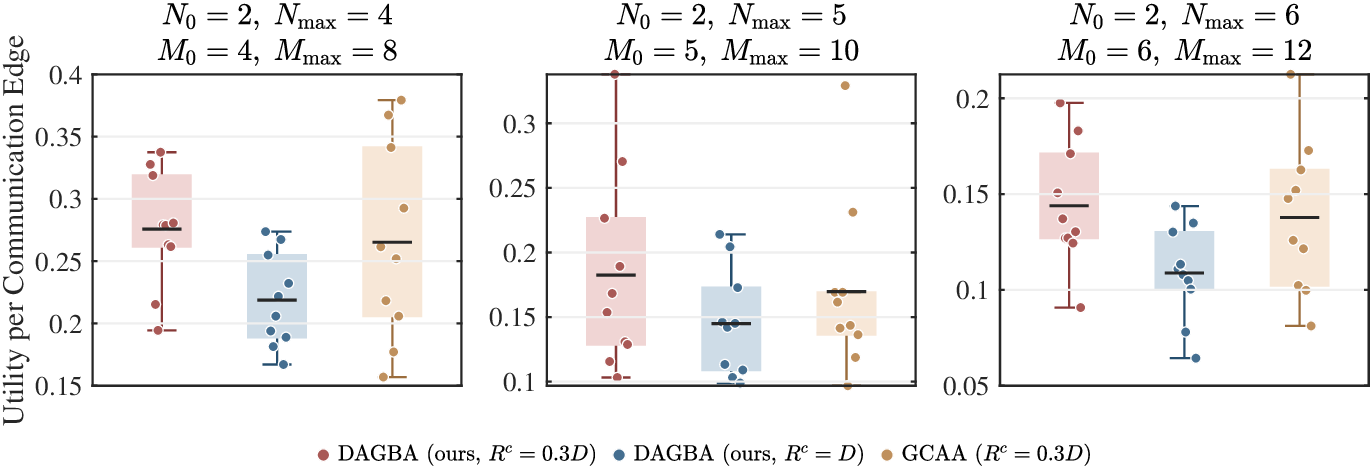}
    \caption{Communication efficiency across problem scales. The metric is computed as the workload-normalized average utility per communication edge. Each box shows the distribution over $10$ independent trials, and dots denote individual trials.}
    \label{fig:algorithm_comm_efficiency}
\end{figure}

\begin{table}[t]
\centering
\caption{Communication efficiency measured by utility per communication edge. Each entry reports mean $\pm$ SE over $10$ independent runs.}
\label{tab:communication_efficiency}
\scriptsize
\setlength{\tabcolsep}{3.0pt}
\begin{tabular}{cccc}
\toprule
Scale
& \begin{tabular}[c]{@{}c@{}}DAGBA\\$(R^c=0.3D)$\end{tabular}
& \begin{tabular}[c]{@{}c@{}}DAGBA\\$(R^c=D)$\end{tabular}
& \begin{tabular}[c]{@{}c@{}}GCAA\\$(R^c=0.3D)$\end{tabular} \\
\midrule
$4/8$  & $0.276 \pm 0.015$ & $0.219 \pm 0.012$ & $0.265 \pm 0.025$ \\
$5/10$ & $0.182 \pm 0.024$ & $0.145 \pm 0.013$ & $0.170 \pm 0.021$ \\
$6/12$ & $0.144 \pm 0.010$ & $0.109 \pm 0.008$ & $0.138 \pm 0.012$ \\
\bottomrule
\end{tabular}
\end{table}

\subsection{Baselines and Metrics}\label{subsec5.2}

\begin{figure}[t]
    \centering
    \includegraphics[width=\linewidth]{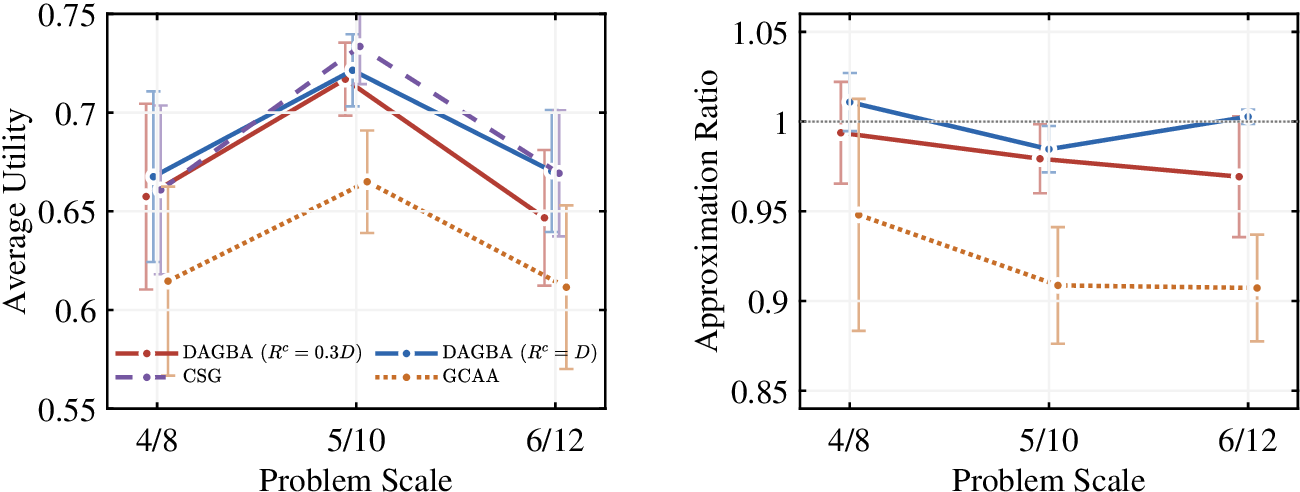}
    \caption{Average utility and empirical approximation ratio relative to CSG across problem scales. Error bars denote the standard error over $10$ independent trials.}
    \label{fig:normalized_utility_scale_trends}
\end{figure}

\begin{figure}[t]
    \centering
    \includegraphics[width=\linewidth]{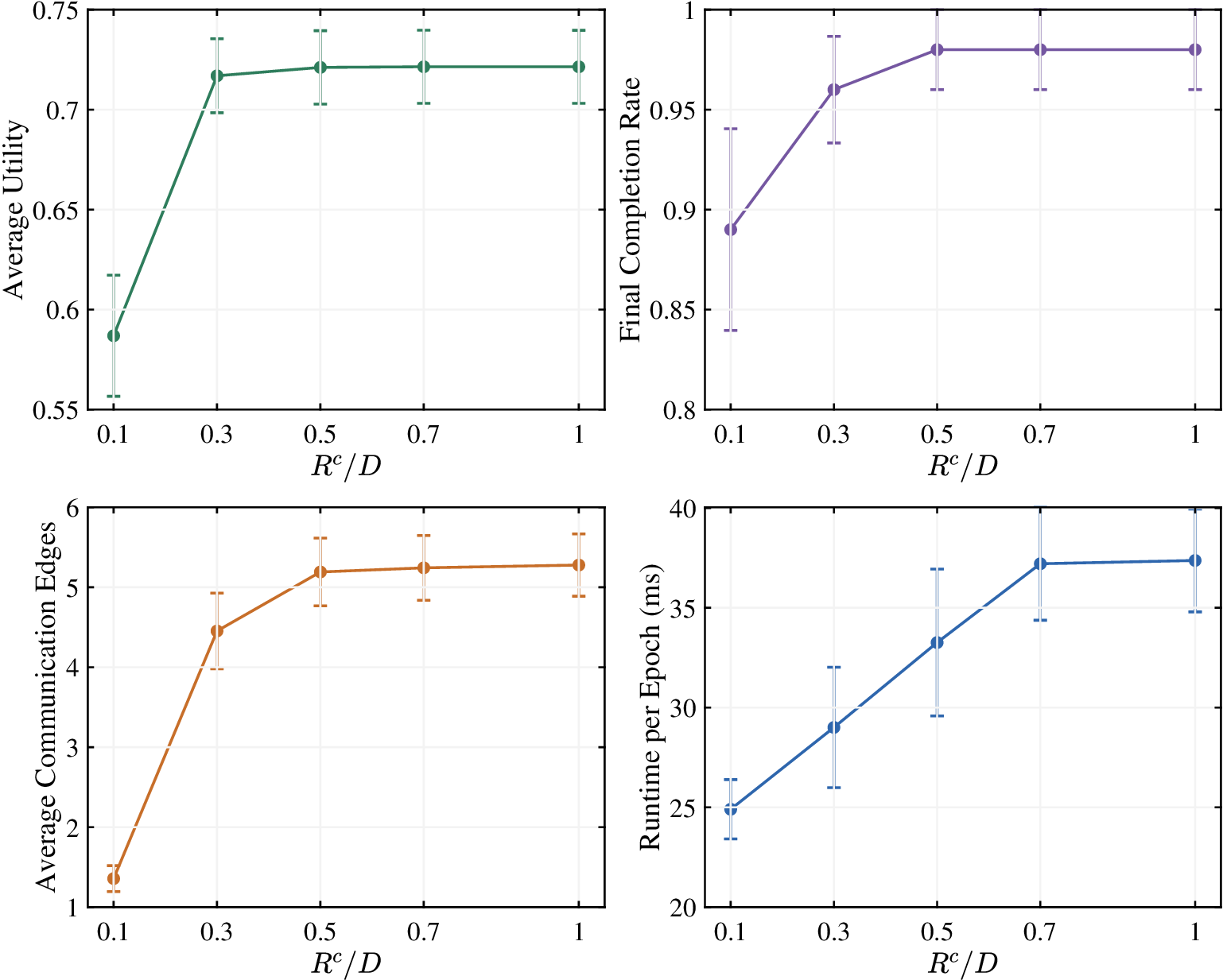}
    \caption{Communication-performance tradeoff under different communication radii. Performance retention and reduction are measured relative to complete communication $R^c=D$.}
    \label{fig:comm_sensitivity_tradeoff}
\end{figure}

\begin{figure}[t]
    \centering
    \includegraphics[width=\linewidth]{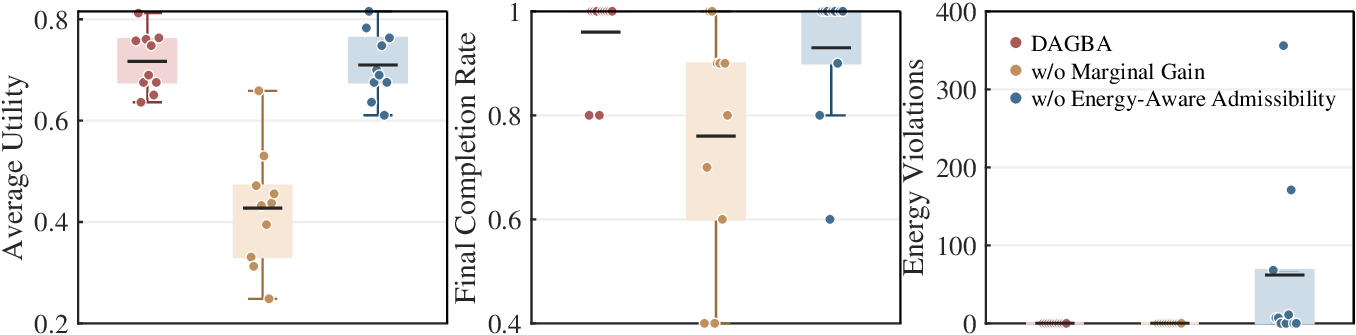}
\caption{Ablation study under sparse communication on the $5/10$ setting. The compared variants remove the submodular marginal-gain rule or the energy-aware admissible-set construction.}
    \label{fig:ablation_summary_core}
\end{figure}

\begin{table*}[t]
\centering
\caption{Communication sensitivity of DAGBA on the $5/10$ setting. Values are reported as mean $\pm$ SE over $10$ independent runs. Reductions are computed relative to the complete-communication case $R^c=D$.}
\label{tab:comm_sensitivity_revised}
\scriptsize
\begin{tabular}{ccccccc}
\toprule
$R^c/D$
& Avg. utility $\uparrow$
& Final completion rate $\uparrow$
& Comm. edges $\downarrow$
& Runtime/epoch (ms) $\downarrow$
& Edge reduction $\uparrow$
& Runtime reduction $\uparrow$ \\
\midrule
$0.1$ & $0.587 \pm 0.030$ & $0.890 \pm 0.050$ & $1.36 \pm 0.16$ & $24.90 \pm 1.48$ & $74.3\%$ & $33.4\%$ \\
$0.3$ & $0.717 \pm 0.019$ & $0.960 \pm 0.027$ & $4.45 \pm 0.47$ & $29.00 \pm 3.02$ & $15.6\%$ & $22.4\%$ \\
$0.5$ & $0.721 \pm 0.018$ & $0.980 \pm 0.020$ & $5.19 \pm 0.42$ & $33.26 \pm 3.68$ & $1.6\%$ & $11.0\%$ \\
$0.7$ & $0.721 \pm 0.018$ & $0.980 \pm 0.020$ & $5.24 \pm 0.40$ & $37.20 \pm 2.83$ & $0.7\%$ & $0.4\%$ \\
$1.0$ & $0.721 \pm 0.018$ & $0.980 \pm 0.020$ & $5.28 \pm 0.39$ & $37.37 \pm 2.58$ & $0.0\%$ & $0.0\%$ \\
\bottomrule
\end{tabular}
\end{table*}

\begin{table*}[t]
\centering
\caption{Ablation study on the $5/10$ sparse-communication setting. Values are reported as mean $\pm$ SE over $10$ independent runs. Higher is better for average utility and final completion rate, while lower is better for energy violations.}
\label{tab:ablation_core}
\scriptsize
\begin{tabular}{lccc}
\toprule
Variant
& Avg. utility $\uparrow$
& Final completion rate $\uparrow$
& Energy violations $\downarrow$ \\
\midrule
DAGBA
& $0.717 \pm 0.019$
& $0.960 \pm 0.027$
& $0.000 \pm 0.000$ \\
w/o Marginal Gain
& $0.427 \pm 0.037$
& $0.760 \pm 0.072$
& $0.000 \pm 0.000$ \\
w/o Energy-Aware Admissibility
& $0.710 \pm 0.021$
& $0.930 \pm 0.042$
& $62.000 \pm 36.875$ \\
\bottomrule
\end{tabular}
\end{table*}

The compared methods include DAGBA under sparse communication, DAGBA under complete communication, centralized sequential greedy assignment (CSG) \cite{fisher1978analysis}, greedy coalition auction algorithm (GCAA) \cite{Braquet2021GreedyDA}, and an energy-feasible weighted random heuristic. DAGBA under sparse communication uses $R^c=0.3D$ unless otherwise specified. DAGBA under complete communication uses $R^c=D$ and serves as a communication-unconstrained implementation of the same distributed marginal-greedy mechanism. CSG is a centralized sequential marginal-greedy baseline evaluated on the same monotone submodular stage utility. GCAA is a distributed auction-type baseline. The random baseline samples from energy-feasible candidate pairs using utility-proportional weights.

We evaluate three problem scales: $N_0=2$, $N_{\max}=4$, $M_0=4$, $M_{\max}=8$; $N_0=2$, $N_{\max}=5$, $M_0=5$, $M_{\max}=10$; and $N_0=2$, $N_{\max}=6$, $M_0=6$, $M_{\max}=12$, denoted by the $4/8$, $5/10$, and $6/12$ settings, respectively. To evaluate task allocation under time-varying agents, tasks, communication links, and residual-energy constraints, we report the following metrics:
\begin{itemize}
   \item The \emph{average utility} of a run is computed as the time average of the workload-normalized stage utility,
   \begin{equation}
    \bar{U}
    =
    \frac{1}{|\mathcal{T}_r^{+}|}
    \sum_{t\in\mathcal{T}_r^{+}}
    \frac{
    U_t(\Pi_t)
    }{
    \sum_{j\in\mathcal{M}_t^{\mathrm{unf}}} \omega_j
    },
    \label{eq:average_utility_metric}
\end{equation}
where $\mathcal{T}_r^{+} = \left\{ t: \sum_{j\in\mathcal{M}_t^{\mathrm{unf}}} \omega_j >0
  \right\}$, $\mathcal{M}_t^{\mathrm{unf}}$ is the set of active unfinished tasks at epoch $t$. This normalization accounts for the time-varying active task load.

	\item  The \emph{task completion rate} at time $t$ is the fraction of appeared tasks completed by time $t$, defined as
\begin{equation}
\mathrm{TCR}(\tau)
=
\frac{
\left|
\left\{
j:
\tau_j^{\mathrm{arr}}\le\tau,\
\tau_j^{\mathrm{com}}\le\tau
\right\}
\right|
}{
\left|
\left\{
j:
\tau_j^{\mathrm{arr}}\le\tau
\right\}
\right|
},
\label{eq:task_completion_rate}
\end{equation}
where $\tau_j^{\mathrm{arr}}$ and $\tau_j^{\mathrm{com}}$ denote the physical arrival and completion times of task $j$, respectively.
The final completion rate is $\mathrm{TCR}(t_f)$ at $t_f=30\,\mathrm{s}$.

\item For each problem scale, the empirical \emph{approximation ratio} relative to CSG is defined as
\begin{equation}
    \mathrm{Ratio}^{\mathrm{alg}}
    =
    \frac{
    \bar{U}^{\mathrm{alg}}
    }{
    \bar{U}^{\mathrm{CSG}}+\epsilon
    },
    \label{eq:csg_normalized_ratio}
\end{equation}
where $\epsilon>0$ is a small numerical constant used only to avoid division by zero. This metric is an empirical utility ratio relative to CSG.

\item The number of \emph{communication edges} is $|\mathcal{E}_t|$, averaged over time and trials. The \emph{per-epoch runtime} is the average allocation runtime per decision epoch. The \emph{communication efficiency} is measured by the workload-normalized average utility per communication edge. Finally, \emph{energy violations} count assignments that would violate residual-energy feasibility when energy-aware admissible-set construction is removed.
\end{itemize}

\subsection{Results and Discussion}\label{subsec5.3}

Figure \ref{fig:revised_main_performance} summarizes the performance under the three problem scales. Figure \ref{fig:mission_average_utility} reports the distribution of the normalized average utility over $10$ independent trials. DAGBA with sparse communication achieves utility distributions close to DAGBA with complete communication and CSG, while generally outperforming GCAA. This indicates that the proposed local marginal-gain bidding rule preserves high-quality stage allocation under limited communication and provides a utility advantage over the auction-type distributed baseline in the tested scenarios. Figure \ref{fig:mission_completion} shows the corresponding task-completion process. DAGBA under sparse communication follows completion trajectories close to DAGBA with complete communication, CSG, and GCAA, whereas the random baseline progresses substantially more slowly. Hence, energy-feasible random selection alone is insufficient for efficient mission execution. {Marginal-gain-based coordination improves both normalized stage utility and finite-horizon task-completion progress.}

Figure \ref{fig:algorithm_comm_efficiency} and \cref{tab:communication_efficiency} compare the communication efficiency of DAGBA and GCAA across the three problem scales. The metric measures the workload-normalized average utility obtained per communication edge and therefore captures the utility gained from each unit of communication connectivity. DAGBA under sparse communication achieves higher communication efficiency than its complete-communication counterpart, showing that increasing the communication radius improves information availability but also introduces additional communication edges that do not always yield proportional utility gains. Compared with GCAA, sparse DAGBA attains competitive or higher utility per communication edge across the tested scales. This indicates that the proposed local marginal-gain bidding mechanism uses limited communication more effectively {and does not rely on dense connectivity to obtain high allocation utility}.

Figure \ref{fig:normalized_utility_scale_trends} compares the average utility and the empirical utility ratio relative to CSG across different problem scales. The average utility directly measures the normalized stage allocation quality, whereas the empirical utility ratio reports the fraction of CSG utility recovered by each method under the same scale. A value close to one indicates that the method attains utility comparable to the centralized sequential greedy reference. DAGBA remains close to the CSG reference across the tested scales and generally achieves a higher ratio than GCAA.

Figure \ref{fig:comm_sensitivity_tradeoff} and \cref{tab:comm_sensitivity_revised} quantify the effect of communication sparsification. Reducing $R^c$ substantially decreases the number of communication edges and the per-epoch allocation runtime. For moderate sparse communication, average utility and final completion rate remain close to the complete-communication reference.

Figure \ref{fig:ablation_summary_core} and \cref{tab:ablation_core} isolate the two main algorithmic ingredients. Removing the submodular marginal-gain rule substantially reduces average utility and final completion rate, showing that DAGBA is not merely a generic bidding protocol {but relies on the greedy marginal-selection principle underlying \cref{thm:sequential_greedy} and~\cref{thm:local_information_approximation}}. Removing the energy-aware admissible-set construction can keep utility high in some runs, but it produces energy violations and therefore infeasible assignments. {This result is consistent with the energy-safety guarantee in  \cref{prop:feasibility_termination}. When the admissibility filter $\widehat{C}_{ij,t}\le E_{i,t}$ is enforced, all returned assignments satisfy $E_{i,t+1}\ge 0$. When it is
removed, this guarantee is violated.} Together, the full DAGBA combines marginal-gain-based coordination for high-quality assignments with energy-aware admissible-set construction for execution feasibility.


\section{Conclusion}\label{sec.6}

{This paper developed DAGBA for energy-constrained distributed task allocation under time-varying agent sets, task sets, and communication graphs. We established per-epoch feasibility, residual-energy safety, finite-round convergence, and approximation guarantees for both the sequential-greedy and local-information regimes. Simulation results demonstrated that DAGBA achieves competitive allocation utility and task-completion performance under sparse communication, while the ablation study verified the complementary roles of marginal-gain bidding and energy-aware admissible-set construction. Future work will investigate non-myopic energy planning and distributed task allocation under more general submodular objectives and communication constraints.}

\bibliographystyle{IEEEtran}
\bibliography{refs}

\end{document}